\documentclass{webofc}
\usepackage[varg]{txfonts}   
\begin{document}

\vspace{-0.5cm}
\noindent
Contribution presented at the 9th International Workshop on Charm Physics, Budker INP, Novosibirsk, Russia, May 21 to 25, 2018.

\title{Parametrizations of three-body hadronic  $B$- and  $D$-decay amplitudes}
%
%

\author{\firstname{Beno\^it} \lastname{Loiseau}\inst{1}\fnsep\thanks{\email{loiseau@lpnhe.in2p3.fr} }\and
        \firstname{Diogo} \lastname{Boito}\inst{2}
         \and
       \firstname{Jean-Pierre} \lastname{Dedonder}\inst{1}
       \and
      \firstname{Bruno} \lastname{El-Bennich}\inst{3}
      \and
       \firstname{Rafel} \lastname{Escribano}\inst{4}
       \and
        \firstname{Robert} \lastname{Kami\'nski}\inst{5}
         \and
       \firstname{Leonard} \lastname{Le\'sniak}\inst{5}
}

\institute{ Sorbonne Universit\'e, Campus Pierre et Marie Curie, Sorbonne Paris Cit\'e, Universit\'e Paris Diderot, et IN2P3-CNRS, LPNHE, Groupe Ph\'enom\'enologie, Paris, France
\and
          Instituto de F\'isica de S\~ao Carlos, Universidade de S\~ao Paulo, SP, Brazil 
\and
           Laborat\'orio de F\'isica Te\'orica e Computacional, Universidade Cruzeiro do Sul, S\~ao Paulo, SP, Brazil
\and
Grup de F\'isica Te\`orica and IFAE, Universitat Aut\`onoma de Barcelona, Bellaterra (Barcelona), Spain
\and
H. Niewodnicza\'nski Institute of Nuclear Physics, PAN, Krak\'ow, Poland
}

\abstract{ A short review of our recent work on amplitude
parametrizations of three-body hadronic weak $B$ and $D$ decays is presented.
  The final states are here composed of three light mesons, namely the various charge
  $\pi\pi\pi$, $K\pi\pi$ and $KK\bar K$ states. 
These parametrizations are derived from previous calculations based on a quasi-two-body
factorization approach where the two-body hadronic final state
interactions are fully taken into account in terms of unitary $S$- and
$P$-wave $\pi\pi$, $\pi K$ and $K \bar K$ form factors.
They are an alternative to the isobar-model description  and can be useful in the 
 interpretation of CP asymmetries.
}
\maketitle
\section{Introduction}
\label{intro}
\subsection{Motivations: why study three-body hadronic  $B$ and  $D$ decays?}
\label{Motivations}

Three-body hadronic  $B$ and  $D$ decays provide a rich tool to study not only the Standard Model, QCD, CP violation~\cite{R.Aaij_LHCbPRD90} but also hadron physics.
The hadron physics, often characterized by two-body resonances and their interferences, affect weak observables and any reliable determination of the later will require a good knowledge of the final state meson-meson interactions. 
This can be realized by introducing theoretical constraints such as  unitarity, analyticity, chiral symmetry and the use of data from reactions other than $B$ and $D$ decays.
Basic Dalitz-plot analyzes rely on sums of relativistic Breit-Wigner amplitudes representing the different possible implied resonances to which some non resonant background amplitude is added.
The $S$-wave resonance  contributions are often difficult to fit.
Can one go  beyond this  isobar model approach?

One can replace the sums of relativistic Breit-Wigner components by parametrizations~\cite{PRD96p113003} in terms of unitary two-meson form factors keeping the weak-interaction dynamics governing the flavor-changing process via $W$-meson exchange.
These parametrizations are based on published results and motivated by analyzes of high-statistics present and forthcoming data at BES III, LHCb, Belle II, Super c-tau factory ....
Up to now there is no three-body decay factorization theorem but major contributions arise from intermediate resonances such as  $\rho(770)$, $K^*(892)$, $\phi(1020)$ which allows to describe three-body decays as quasi-two-body ones.
 For instance, for the  three-meson final state of the $D^0 \to K^0_S \pi^- \pi^+ $ decay,  one can introduce quasi-two-body pairs, $[K^0_S \pi^+]_{L} \ \pi^-$, $ [K^0_S \pi^-]_{L} \ \pi^+$, $K^0_S \  [\pi^+{\pi^-}]_{L}$,  two of the three mesons forming a state of angular momentum $0$ or $1$ with $L=S$ or $P$, respectively.
 
 \subsection{QCD quasi-two-body factorization}
 \label{qcdfactorization}
 
 Decays are mediated by local four-quark operators $O_i (\mu)$ forming the weak effective nonrenormalizable Hamiltonian $\mathcal{H}_\mathrm{eff}$.  Schematically for $  B \to M_1 M_2^* (M_2^* \to M_3 M_4)$ one has
 
\begin{equation}
   \langle M_1 M_2^* |  \mathcal{H}_\mathrm{eff} |  B\rangle = \frac{G_F}{\sqrt{2}}\ V_\mathrm{CKM} \sum_i  C_i(\mu)  \langle M_1M_2^* | O_i(\mu) | B\rangle,  
 \label{fullamphamilton}
\end{equation}
where $G_F$ is the Fermi decay constant,  $V_\mathrm{CKM}$ the product of Cabibbo-Kobayashi-Maskawa matrix elements and $C_i(\mu)$ Wilson coefficients renormalized at scale $\mu\sim m_b$ (or $m_c$ in $D$ decays).
In the factorization approach~\cite{Buchalla:1995vs}  with the strong coupling $\alpha_s(\mu)$, i.e. at  scale $\mu$,

 \begin{align}
 \langle M_1 M_2^* |  O_i(\mu) | B \rangle  & =   \Big ( \langle M_1  |  {J}_{1}^{\nu} | B \rangle  \langle M_2^* | J_{2\nu} | 0 \rangle   \nonumber \\
    +  \langle M_1 |  J_{3}^{\nu} | 0   &  \rangle  \langle M_2^* | J_{4\nu} | B  \rangle \Big ) 
    \left [ 1 + \sum_n r_n \alpha_s^n(\mu) + \mathcal{O}  \left ( \frac{\Lambda_\mathrm{QCD}}{m_b} \right ) \right ], 
 \label{QCDfatorization}
\end{align}
 where $r_n$ are strong interaction constant factors and $ | 0 \rangle $ the vacuum state.
For the leading order the factorization takes place with either weak quark currents $J_1$, $J_2$ or $J_3$, $J_4$.
The radiative corrections can be evaluatd to a given order $\alpha_s^n (\mu)$.
The nonperturbative corrections to the heavy-quark limit $ \mathcal{O}  \left ( \frac{\Lambda_\mathrm{QCD}}{m_b} \right )$ are less reliable for $D$ decays as $m_c \sim m_b/3$;  therefore, even though the factorization is more phenomenological for charmed mesons, it can still represent a good starting point.
 
  The amplitude $\langle M_1  | J_{1}^\nu | B \rangle $ ($=\langle M_1  \overline {B}| J_{1}^\nu | 0 \rangle$)  is a heavy-to-light transition form factor which can be evaluated within light-front and relativistic constituent quark models, light-cone sum rules, continuum functional QCD and lattice QCD (see Appendix A4 of Ref.~\cite{PRD96p113003}).
Semi-leptonic decay measurements like $D^0 \to \pi^- e^+ \nu_e$ can also allow a phenomenological determination of these form factors.

The matrix element   $\langle M_2^* | J_{2\nu} | 0 \rangle \propto \langle M_3 M_4 | J_{2\nu} | 0 \rangle$, where the $M_3 M_4$ resonance, $M_2^*$, originates  from a $\bar q q$ pair, corresponds to the $M_3 M_4$ form factor.
It has been shown, in  Ref.~\cite{Barton65}, that, using dispersion relations and field theory,  this form factor can be fully determined, if the $M_3M_4$ strong interaction is known at all energies.
These form factors are calculated  from Muskhelishvili-Omn\`es equations~\cite{MO} using two-body  data, unitarity, asymptotic QCD and chiral symmetry constraints at low energies.

The term  $\langle M_1 | J_{3}^\nu | 0 \rangle$, related to the $M_1$ weak decay constant, is known from experiment, e.g. the pion decay constant, $f_\pi$ or that of the kaon, $f_K$.
It can also be evaluated with lattice-regularized QCD and other nonperturbative approaches.

The matrix element $\langle M_2^* | J_{4\nu} | B \rangle \propto \langle M_3 M_4 | J_{4\nu} | B \rangle$ corresponding to $B$ meson transitions to  two-meson pairs  via the $M_2^*$ resonance is the biggest uncertainty in our approach.
It could be evaluated from semi-leptonic processes: like $B^0 \to K^+ \pi^- \mu^+ \mu^-$ or $D^0 \to K^- \pi^+ \mu^+ \mu^-$.
 In the derivation of the amplitude  presented here it will be related to the $\langle M_2^*[\to M_3 M_4]| {J}_{2\nu} | 0 \rangle $ form factor.
Within the soft-collinear effective theory, the amplitude can be factorized in terms of generalized $B$-to-two-body form factor and two-hadron light-cone distribution amplitude~\cite{KleinJHEP10}.

\subsection{Application to the $D^+ \to K^- \pi^+ \pi^+ $ decay}
\label{Dkpipi}

In this process, studied in Ref.~\cite{PRD80_054007}, the final state $\pi^+ \pi^+$ interaction can be neglected and the quasi-two-body $[ K^- \pi^+]_{S,P}\  \pi^+ $ can be introduced.
There is no penguin contribution (loop with $W$ meson) and only the effective Wilson-coefficients $a_{1(2)}$ appear in the quasi-two-body factorized amplitude,

\begin{align}
\label{Dkpipifact}
\langle [ K^- \pi^+]_{S,P}\  &\pi^+ | \mathcal{H}_\mathrm{eff} | D^+ \rangle =  \frac{G_F}{\sqrt{2}}\cos^2\theta_C 
\Big [a_1\langle [K^-\pi^+_1]_{S,P}\vert \bar s\gamma^\nu(1-\gamma_5)c\vert D^+\rangle  \langle\pi^+_2 \vert \bar u\gamma_\nu(1-\gamma_5)d\vert 0\rangle    \nonumber    \\
                                               &+ a_2  \langle [K^-\pi^+_1]_{S,P}  \vert \bar s\gamma^\nu(1-\gamma_5)d \vert 0\rangle
                                                    \langle \pi^+_2 \vert \bar u\gamma_\nu(1-\gamma_5)c \vert D^+\rangle \Big ] +(\pi_1^+ \leftrightarrow \pi_2^+), 
\end{align}
$\theta_C$ being the Cabbibo angle.
The matrix element  $\langle[K^-\pi^+_1]_{S,P} \vert \bar s\gamma^\nu(1-\gamma_5)d \vert0\rangle$ is given by the $K\pi$ form factors.
The term $\langle [K^-\pi^+_1]_{S,P}\vert \bar s\gamma^\nu(1-\gamma_5)c \vert D^+\rangle$ is less straightforward to evaluate.
Assuming a dominant intermediate resonance $R$, it can be written as being proportional to the $D$ to $R$ $[R \to K \pi]$ transition form factor multiplied by the $K\pi$ form factors.
This description is a feature of crucial importance to our proposed parametrizations. 
In Eq.~(\ref{Dkpipifact}), $ \langle\pi^+_2(p) \vert \bar u\gamma_\nu(1-\gamma_5)d\vert 0\rangle=-if_\pi p_\nu$ and $\langle \pi^+_2 \vert \bar u\gamma_\nu(1-\gamma_5)c \vert D^+\rangle$ is the $D \pi$ transition form factor.
 
 Parametrized amplitudes based on  quasi-two-body factorization have been given in Ref.~\cite{PRD96p113003} in terms of analytic and unitary meson-meson form factors  for final states composed of three light mesons, namely the various charge $\pi\pi\pi$, $K\pi\pi$ and $KK\bar K$ states.
For these hadronic three-body decays we have shown, in previous studies, that this approach is phenomenologically successful.
Below, we illustrate these parametrizations for the $B \to K \pi^+\pi^-$~\cite{fkll, ElBennich2006, ElBennichetal09} and   $D^0 \to K^0_S K^+\ K^-$~\cite{JPD_inprogress} for  meson-meson final states in  $S$ wave.
Formulae for meson-meson final states in $P$ wave are given in Ref.~\cite{PRD96p113003}.

\section{Parametrized amplitudes for the $ B \to K \pi^+ \pi^-$ decays}
\label{paraBKpipi}
 
\subsection{Parametrization of  the $B \to K [\pi^\pm \pi^\mp]_{S}$ amplitudes}
\label{paraBKpipi_S}

Let us label the momenta as  $B(p_B)\to K(p_1) \pi^+(p_2)\pi^ -(p_3)$ with $s_{12}=(p_1+p_2)^ 2$, $s_{13 }=(p_1+p_3)^ 2$,  $s_{23}=(p_2+p_3)^ 2$ and $s_{12}+ s_{13}+s_{23}= m_B^2+m_K^2+2 m_\pi^2$ .
As can be seen from Eq.~(1) of Ref.~\cite{fkll}  the $B \to K [\pi^+ \pi^-]_{S}$ amplitude can be parametrized in terms of three complex parameters, $ b_i^S, i=1,2,3$, for the different charged states $B= B^\pm, K =K^\pm$ and $B= B^0 (\bar B^0), K = K^0(\bar K^0)$ or $K^0_S$. For the $B^-$ decays one has

\begin{align}
  \mathcal{A}_S(s_{23})  &\equiv  \langle K^- \ [\pi^+\pi^-]_S \vert \mathcal{H}_{\rm eff} \vert B^- \rangle 
 =  b_1^S \left (M_B^2-s_{23} \right ) F_{0n}^{\pi\pi} (s_{23}) + \left ( b_2^S F_0^{B K}(s_{23})+b_3^S \right ) F_{0s}^{\pi\pi}  (s_{23}),
\end{align}
where $F_0^{B K}(s)$ is the $B$ to $K$ transition form factor (see Refs.~\cite{PRD96p113003, fkll}). 
The non-strange scalar form factor $F_{0n}^{\pi \pi}(s)$ contains the contributions of $f_0(500$), $f_0(980)$ and $f_0(1400)$.
Several models are compared in Fig.~8 of Ref.~\cite{JPD_PRD89}.
Although there are large differences, it has been checked by the authors that, with the fitted form factor to obtain the lowest $\chi^2$ for $D^0 \to K^0_S \pi^+\pi^-$, the main conclusions achieved for the $B^\pm \to  \pi^+\pi^-\pi^\pm$ in Ref.~\cite{Dedonderetal2011} were unchanged (see Ref.~\cite{JPD_PRD89} for explanations).
The modulus of the Moussallam pion scalar form factor~\cite{Moussallam_2000}, calculated by solving the Muskhelishvili-Omn\`es equation~\cite{MO}, is close to that of the form factor obtained in Ref.~\cite{JPD_PRD89}, notably below 1 GeV.
 A plot of the strange scalar form factor $F_{0s}^{\pi \pi}(s)$, which receives the contribution of the $f_0(980)$ and $f_0(1400)$,  can be found in Fig.~6 of Ref.~\cite{EPJC78_1000}.
 It has been calculated using the Muskhelishvili-Omn\`es approach.

 In terms of the original amplitude~\cite{fkll} one has\footnote{The interested reader will find, in Appendix B of Ref.~\cite{PRD96p113003}, the corresponding relations for the other parameters.}, $F_0^{B\to f_0(980)}(m_K^2)$ being the $B$ to $f_0(980)$ transition form factor evaluated at $m_K^2$~\cite{PRD96p113003},

\begin{equation}
 b_{1}^{S}= \frac{G_F}{\sqrt{2}}\left[\chi f_K F_0^{B\to f_0(980)}(m_K^2)\ U
-\tilde C\right],
\end{equation}
where $\tilde C=f_\pi F_\pi\left(\lambda_uP_1^{GIM}\!\!+\!\lambda_tP_1\right)$, $\lambda_u= V_{ub} V_{us}^*$, 
 $ \lambda_t= V_{tb} V_{ts}^*$, $F_\pi$ is the $B\pi$ form factor at $m_\pi^2=0$, $P_1^{GIM}$, $P_1$ complex charming penguin parameters and $U$ is a short-distance contribution given in terms of CKM matrix element multiplied by effective Wilson coefficients.
 The fitted parameter $\chi$ represents the strength of the non-strange pion form factor contribution, furthermore.
 Its value can be estimated from the $f_0(980)$ decay properties~\cite{fkll}.
A summary of the models for the scalar-isoscalar pion form factor can be found in Appendix A4 of Ref.~\cite{PRD96p113003} and, as just noted above, see also the recent determination of Ref.~\cite{EPJC78_1000} and the review talk~\cite{EP2019MITP}.

\subsection{Parametrization of  the $B \to [K \pi^\pm]_{S}\pi^\mp$ amplitudes}
\label{paraBKpi_Spi}

In terms of the two complex parameters  $c_1^S$, $c_2^S$ (see  Eq.~(68) of Ref.~\cite{ElBennichetal09}) one has

\begin{equation}
\mathcal{A}_S (s_{12}) \equiv \langle \pi^- \ [ K^- \pi^+]_S \vert \mathcal{H}_\mathrm{eff}\vert B^- \rangle = 
    \big ( c_1^S + c_2^S s_{12} \big )\, \frac { F_0^{ B \pi } (s_{12} ) F_0^{K \pi} (s_{12} )}{s_{12}}, 
\end{equation}
where $F_{0}^{K \pi } (s)$ (contribution of $K^*_0(800)$ or $\kappa$ and of $K^*_0(1430)$, see e.g. Fig.~7 of Ref.~\cite{JPD_PRD89}) and $F_{0}^{ B\pi}(s)$ are the $K\pi$ and $B\pi$ scalar  form factors, respectively.
This parametrization  has been used with success in the amplitude analysis~\cite{BK0SppLHCb17} of the Dalitz-plot distribution of the LHCb $\bar B \to K^0_S \pi^+ \pi^-$  data.
One has~\cite{ElBennichetal09}
 
  \begin{equation}
c_1^{S}=\dfrac{G_F}{\sqrt{2}} (M_B^2-m_{\pi}^2)(m_{K}^2-m_{\pi}^2)
\left[ \lambda_u\left(a_4^{u}(S)-\dfrac{a_{10}^{u}(S)}{2}+c_4^{u}\right)
+\lambda_c\left(a_4^c(S)-\dfrac{a_{10}^c(S)}{2}+c_4^{c}\right)\right],
\end{equation}
where $\lambda_c= V_{cb} V_{cs}^*$.
The  $a_i^{u(c)}(S)$, $i=4, 10$ are the  leading order effective Wilson coefficients  including vertex and penguin corrections.
The $c_4^{u(c)}$ are free fitted parameters simulating non-perturbative  and higher order contributions to the penguin diagrams.
Models for the  $F_{0}^{K \pi}(s)$ form factor are described in Ref.~\cite{PRD96p113003}, see also some complementary aspects in Ref.~\cite{EP2019MITP}.

\section{$D^0 \to K^0_S [K^+ K^-]_S$ and $D^0 \to [K^0_S K^\pm]_S K^\mp$ parametrized amplitudes} 
\label{D0K0SKKparam}

The $[K^+ K^-]$ pairs can have isospin 0 or 1 but the $[K^0_S K^\pm]$ ones have isospin 1.
The $f_0(980)$, $f_0(1400)$, $a_0(980)^0$ and $a_0(1450)^0$  contribute to the following parametrized  amplitude

\begin{equation}
\label{AS00}
{\mathcal{A}}^0_{S,0}(s_{23}) =  {{h}_1^S \left(m_{D^0}^2-s_{23} \right) F^{ K\bar K}_{0n} (s_{23} ) + h_2^S \left(m_{K^0}^2-s_{23} \right) F^{ K\bar K}_{0s} (s_{23} )  +
{h}_3^S \left(m_{D^0}^2-s_{23} \right)G_0^{K \bar K}(s_{23})}.
\end{equation}
 where $s_{23}$ is the energy  squared of the $K^+K^-$ pair while $s_{12}$ is associated to the $K^0_S K^-$ pair and $s_{13}$ to the $K^0_S K^+$ one.
The decay amplitude  associated with the $a_0(980)^-$ and $a_0(1450)^-$ resonances, can be parametrized as: 

 \begin{equation}
 \mathcal{A}_{S,-}^0(s_{12})=\left(h_4^ S+h_5^ S s_{12}\right)  G_0^{K \bar K}(s_{12}).
 \label{ASM0}
\end{equation}
 The amplitude carrying contributions from  $a_0(980)^+$ and $a_0(1450)^+$] reads
 
 \begin{equation}
 \mathcal{A}_{S,+}^0(s_{13})=\left[h_6^S \dfrac{F_0^{D K}(s_{13})}{s_{13}}+h_7^ S(m_{K}^2-s_{13})\right]   G_0^{K\bar K}(s_{13}).
 \label{ASP0}
 \end{equation}
  Models for the $F^ {K\bar K}_{0n(s)}(s)$ form factors entering Eq.~(\ref{AS00}) have been derived in Ref.~\cite{FurmanPLB699_102, LZ2014} (see their Figs.~1) solving three coupled   channels viz. $\pi \pi$, $K \bar K$ and $4\pi$ (effective $2\pi$-$2\pi$ or $\sigma \sigma$ or $\rho \rho$ ...) and imposing chiral symmetry constraints.
  The $F^ {K\bar K}_{0s}(s)$   form factor has also been calculated in a dispersive approach in Ref.~\cite{EPJC78_1000} (see  their Fig.~7).
  
In Eqs.~(\ref{ASM0}) and~(\ref{ASP0}),  the scalar-isovector $G^{K\bar K}_0(s)$ form factor, built  in  Ref.~\cite{EPJC_75_488} from a unitary $S$-wave coupled channel ($\eta \pi, K \bar K$) model, is plotted in their Fig.~7.
This model, derived from the Muskhelishvili-Omn\`es equation~\cite{MO}, imposes the presence of the $a_0(980)$ and  $a_0(1450)$ and includes asymptotic QCD and chiral symmetry constraints.
Models for the transition form factor $F_0^{D K}(s)$ in Eq.~(\ref{ASP0}) can be found in Ref.~\cite{PRD96p113003}.
The above complex $h_i^S$ coefficients are given in terms of the original amplitudes  in Appendix B of Ref.~\cite{PRD96p113003}.

\section{Concluding remarks}

  Alternatives to isobar Dalitz-plot model for weak $D$, $B$ decays into various $\pi\pi\pi$, $K\pi\pi$ and $KK\bar K$ charge states have been presented in Ref.~\cite{PRD96p113003}.
 Let us recall that isobar parametrizations do not respect unitarity and extraction of  strong CP phases should be taken with caution. 
 Furthermore $S$-wave resonance contributions are hard to fit.
 
  Our parametrizations,  although not fully three-body unitary,  are based on a sound theoretical application of QCD factorization to a hadronic quasi-two-body decay. 
They assume that  final  three-meson state are preceded by  intermediate resonant states which is justified by phenomenological and experimental evidence.
Analyticity, unitarity, chiral symmetry  plus correct asymptotic behavior of the two-meson
scattering amplitude in $S$ and $P$ waves are implemented via analytical and 
unitary  $S$- and $P$-wave $\pi\pi$, $\pi K$ and $K\bar K$ form factors
entering in hadronic final states of our amplitude parametrizations.

 These parametrized  amplitudes can be readily used adjusting  parameters in a least-square
fit to the Dalitz plot  for a given decay channel and 
employing tabulated form factors as functions of momentum squared or energy.
 The reproduction of the Dalitz-plot data  might require some adjustment of the meson-meson form factors. The addition of phenomenological amplitudes  (contributions of higher interacting waves, in particular $D$ waves or $J$=2 resonances),  and possible three-body rescattering effects may be necessary.

We have exemplified here expressions for the $B \to K \pi^+\pi^-$~\cite{fkll, ElBennich2006, ElBennichetal09} and   $D^0 \to K^0_S K^+\ K^-$~\cite{JPD_inprogress} for  meson-meson final states in  $S$ wave. 
In Ref.~\cite{PRD96p113003} one can find  other explicit amplitude expressions for  meson-meson final states in  $S$ and $P$ wave for  $B^\pm \to \pi^+ \pi^- \pi^\pm$, $B \to K\ \pi^+ \pi^-$, $B^\pm \to K^+ K^-K^\pm$, $D^+ \to \pi^- \pi^+ \pi^+$, $D^+ \to K^- \pi^+ \ \pi^+ $, $D^0 \to K^0_S \ \pi^+\ \pi^- $.
Previous studies have shown that this approach is successful.
In addition, expressions for $D^0 \to K^0_S \ K^+ K^-$  are also given in Ref.~\cite{PRD96p113003}.
 We have derived  preliminary parametrized amplitudes for  the $B^\pm \to K^+ K^- \pi^\pm$ decays~\cite{R.Aaij_LHCbPRD90, CL.Hsu_BellePRD96} and for the $B^0 \to K^0_S \ K^+ K^-$  process presently analyzed by the LHCb collaboration.

\vspace{0.05cm}
\begin{acknowledgement}
B.L. thanks A. Bondar and S. Eidelman for their kind invitation to present this contribution to this CHARM2018 international workshop.
B.E. and B.L. are grateful to the Mainz Institute for Theoretical Physics (MITP) for its hospitality and its partial support during the completion of this work.
\end{acknowledgement}

\end{document}